\documentclass[times, 10pt,twocolumn]{article} 
\usepackage{latex8}
\usepackage{times}
\usepackage{xcolor}
\usepackage{graphicx}
\usepackage{amsmath, amsfonts}
\usepackage{dsfont}
\usepackage{tabularx}
\usepackage{booktabs}

\begin{document}

\title{Towards Robust and Truly Large-Scale Audio--Sheet Music Retrieval}

\author{Lu{\'i}s Carvalho$^1$, Gerhard Widmer$^{1, 2}$ \\
$^1$Institute of Computational Perception \& $^2$LIT Artificial 
Intelligence Lab \\ 
Johannes Kepler University Linz, Austria \\ 
\{luis.carvalho, gerhard.widmer\}@jku.at\\
}

\maketitle
\thispagestyle{empty}

\begin{abstract}

   A range of applications of multi-modal music information retrieval is 
   centred around the problem of connecting large collections of sheet music (images) to 
   corresponding audio recordings, that is, identifying pairs of audio and score excerpts that refer to the same musical content.
   One of the typical and most recent approaches to this task 
   employs cross-modal
   deep learning architectures to learn joint embedding spaces that link
   the two distinct modalities -- audio and sheet music images.
   While there has been steady improvement on this front over the past years, 
   a number of open problems still prevent large-scale employment of this methodology.
   In this article we attempt to provide an insightful examination of the current 
   developments on audio--sheet music retrieval via deep 
   learning methods.
   We first identify a set of main challenges on the road towards robust and 
   large-scale cross-modal music retrieval in real scenarios.
   We then highlight the steps we have taken so far to address some of these 
   challenges, documenting step-by-step improvement along several dimensions.
   We conclude by analysing the remaining 
   challenges and present ideas for solving these, in order to pave the way to a unified and robust methodology for cross-modal music retrieval.
   
\end{abstract}



\section{Task, Basic Approach, and Challenges}

A fundamental paradigm in the field of Music Information Retrieval (MIR) is 
consists in searching and retrieving items of different modalities, 
for example video clips, live and studio recordings, scanned sheet music, 
and album covers.
Moreover, the large amounts of music-related contents that are currently 
available in the digital domain demand for the development of \textit{fast} 
and \textit{robust} 
retrieval methods that allow such extensive and rich collections to be
searched and explored in a content-based way.

\begin{figure*}[htbp]
  \centering
  \includegraphics[width=1\textwidth]{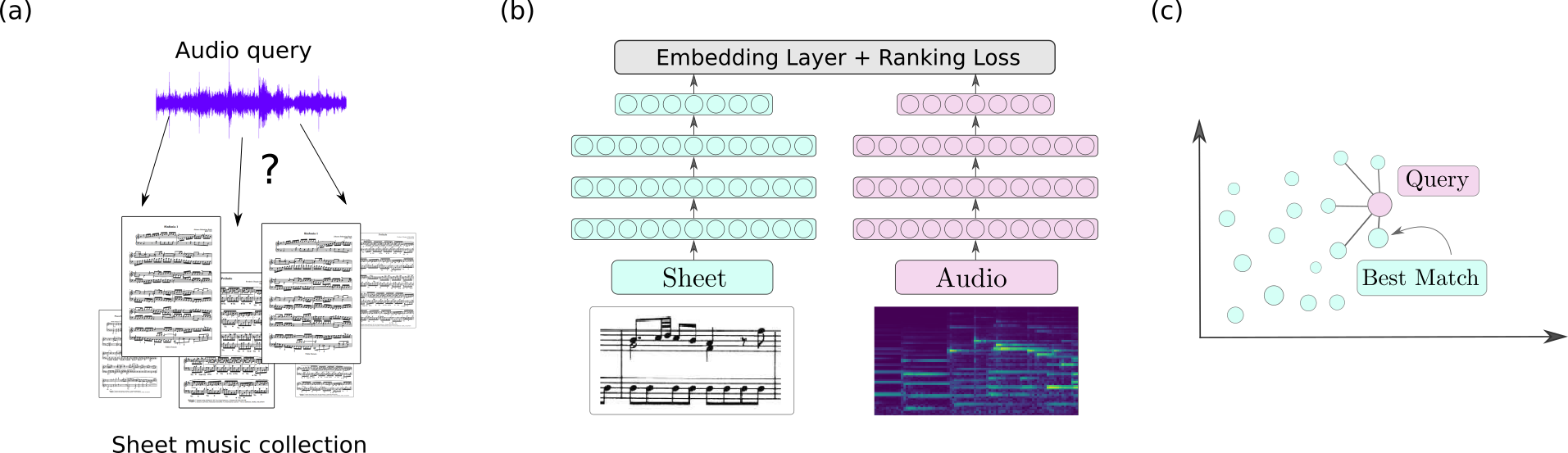}
  \caption{(a) Illustration of the retrieval application. 
  (b) Architecture of the embedding learning network.
  (c) Simplified visualisation of the embedding space.}
\label{fig:main}
\end{figure*}

A central and challenging problem in many cross-modal retrieval scenarios is known 
as \textit{audio--sheet music retrieval}.
The goal here is to, given a query fragment in one of the two 
modalities (a short audio excerpt, for example), retrieve the relevant 
music documents in the counterpart modality (sheet music scans).
In addition, it is typically the case that no 
metadata or machine-readable information (i.e. MIDI or MusicXML formats) 
is available: one has to work directly with raw music material, i.e., 
scanned music sheet images and digitised audio recordings.
Figure~\ref{fig:main}a illustrates the retrieval task when searching an
audio recording within a sheet music collection.

A key step towards audio--sheet music retrieval is to define a convenient
joint representation in which both modalities can be readily 
compared.
The common approaches for defining such shared space rely on handcrafted 
mid-level representations~\cite{MuellerABDW19_MusicRetrieval_IEEE-SPM}, 
such as chroma-based features~\cite{FremereyCME09_SheetMusicID_ISMIR}, 
symbolic fingerprints~\cite{ArztBW12_SymbolicFingerprint_ISMIR}, 
or the bootleg score~\cite{Tsai20_LinkingLakhtoIMSLP_Bootleg_ICASSP}, 
the latter one being a coarse codification of the major 
note-heads of a sheet music image.
However, in order to generate such representations a number of error-prone
pre-processing steps are still needed, i.e., automatic music 
transcription~\cite{SigtiaBD16_DNNPolyPianoTrans_TASLP}
for the audio part, and optical music 
recognition~\cite{Calvo-ZaragozaHP21_OMRReview_ACM} on the sheet music 
side.

A solution avoiding such problematic pre-processing components was proposed 
in~\cite{DorferHAFW18_MSMD_TISMIR}, by designing a deep convolutional network 
(CNN) that can learn an embedding space that is shared between the two 
underlying modalities.
As sketched in Figure~\ref{fig:main}b, this architecture has two 
independent convolutional 
pathways, each being responsible for encoding short fragments of its
respective music modality into a 32-dimensional embedding vector.
This network is fed with pairs of short snippets of sheet music images and 
magnitude spectrograms, and the embedding space is obtained by minimising the 
cosine distance between pairs of matching audio--sheet music snippets, while 
maximising the distance between non-matching pairs.
Training is done by optimising a pairwise ranking loss function, and the 
final canonically correlated layer (CCA)~\cite{DorferSVKW18_CCALayer_IJMIR} 
forces the embeddings computed from 
matching pairs to be correlated to each other in the shared latent space.
Then, when the training is finished, snippet-wise retrieval reduces to nearest neighbour search in the joint space 
(see Figure~\ref{fig:main}c),
which is a simple and fast procedure.
This general retrieval framework based on short segments (snippets) extracted from the larger original documents (audio recordings, complete scores) supports a variety of possible applications, from piece identification to version detection and music recommendation.

The deep learning approach is still in its early stages, and a number of obstacles and open problems prevent robust and large-scale deployment under real-world conditions, some of which we have already begun to solve:
%
\begin{itemize}
    
    \item \textbf{Variable tempo and context discrepancies.}
    Global and local tempo deviations are inherent in performed music and require 
    careful design of the amount of temporal context to be provided
    to a retrieval system during training.
    
    \item \textbf{Strongly-aligned data constraint.}
    Obtaining matching pairs of short excerpts for training deep 
    learning-based models requires finer alignments between audio 
    and sheet music.
    Such data is complex and of expensive nature, and as a result
    synthetic data is used for training.
    
    \item \textbf{Generalisation to real-world (noisy) data} 
    The large numbers of precisely aligned pairs of audio and score snippets required for training are currently obtained by synthesising them in a controlled way from machine-readable scores and corresponding MIDI files. Generalising to real performance recordings and imperfect score scans turns out to be very challenging.
    
    \item \textbf{Temporal dependencies between subsequent snippets.} 
    When handling entire documents, consecutive snippets exhibit strong
    temporal correspondences, which should be exploited for more robust identification and retrieval.
    
    \item \textbf{Public large-scale datasets.}
    Up to this date, there is no license-free and truly large 
    audio--sheet music dataset for evaluation of current algorithms.
    
    \item \textbf{Efficient structures for fast retrieval.}
    Quick cross-modal retrieval algorithms are essential when one is
    browsing through large-scale and heterogeneous music collections.
    This aspect can be often overlooked when the main focus is on 
    retrieval quality metrics such as precision and recall.
    
    
    \item \textbf{Instrumentation and genre.} 
    Current methods have been developed specifically for classical and, even more specifically,
    piano music data. Other types of scores (e.g., orchestral), instruments, and genres will present new complications.
\end{itemize}

In this article, we examine these challenges one by one. 
We first summarise our efforts to address some of the points above, as well 
as the improvements we obtained over the first and original system architecture.
We then turn to the still open problems and propose concrete ideas to address these remaining 
challenges, aiming to establish a unified and robust methodology 
for cross-modal music retrieval in the context of truly large collections of musical materials.

\Section{Some First Solutions}

\subsection{Variable tempo and context discrepancies}\label{sec:temp_dif}

A key limitation of the baseline deep learning solution relates to the 
temporal context (or field of view) that is input to the network: 
both audio and sheet music snippets are fixed in size
(see the inputs of the main model in Figure~\ref{fig:main}b) for a visual
example).
For the audio part, the fragments span roughly 2.1 seconds, which 
corresponds to 42 spectrogram frames.
For the scores, snippets span $160 \times 180$ pixels, after sheet 
music pages being re-scaled to a $1181 \times 835$ resolution.

This implies that the amount of actual musical content within the fragments 
can vary significantly due to the duration of the notes and the tempo in which 
the piece is being performed.
For instance, a sheet music snippet with longer notes played slowly would 
cover a substantially larger duration in the audio than another one with 
shorter notes that has been played faster.
As a consequence, generalisation issues can occur due to 
differences between what the network sees during training and the data it 
will see at application time: 
the musical content fed to the CNN may exhibit considerably less information 
than fragments it has been trained on.

To address this problem, we proposed in~\cite{BalkeDCAW19_ASR_TempoInv_ISMIR}
to let the network learn to adjust the temporal content of a given audio 
excerpt by using a separate \textit{soft-attention mechanism}.
First, the audio excerpt size is considerably expanded, up to four times 
the original duration.
We then append to the audio network an attention pathway which, taking as input 
the audio magnitude spectrogram query, generates a 1-D probability 
density function that has the same number of frames as the input  
spectrogram and acts as an attention mask.
Then, before the spectrogram excerpt is fed into the original audio 
embedding network, each frame thereof is multiplied by its attention 
mask, in this way cancelling out irrelevant parts of the query excerpt and 
focusing on the important information that should be embedded.

In~\cite{BalkeDCAW19_ASR_TempoInv_ISMIR} we conducted a series of quantitative 
and qualitative experiments on synthesised piano music data, with the 
results indicating that the attention mechanism is capable of 
increasing the robustness of the audio--sheet music retrieval system.
Table~\ref{tab:ismir19_results} summarises the main experimental 
results for a snippet-wise 
retrieval scenario: given an audio fragment as a search query, we 
desire to retrieve the matching sheet music snippet within a pool
of 10,000 candidates from the MSMD dataset~\cite{DorferHAFW18_MSMD_TISMIR}.
We compare the baseline network BL1 from~\cite{DorferHAFW18_MSMD_TISMIR} with an 
upgraded version BL2 of it (which replaces the last global pooling layer of each 
modality pathway with a dense layer) and check for retrieval improvements 
when adding the attention mechanism (+AT) 
and increasing the duration of the audio excerpts from a short 
context (SC, 2.1 sec) to a long context (LC, 8.4 sec).

No improvement is at first observed when only expanding the temporal 
context of the second baseline (from BL2-SC to BL2-LC).
However, when appending the attention mechanism to BL2, we notice a 
boost in retrieval performance, with the MRR increasing from 0.63 
to 0.75.
When comparing the main baseline BL1-SC with our best model configuration, 
we observe a substantial improvement in all evaluation metrics 
(MRR increases by 0.44 points).

\begin{table}[!t]
\centering
\scalebox{0.92}{
 \begin{tabular}{lcccccc}
 \toprule
 \textbf{Model} & \bfseries R@1 & \bfseries R@5 & \bfseries R@25 & \bfseries MRR & \bfseries MR\\
 \midrule
 BL1-2s & 19.12 & 44.16 & 66.63 & 0.31 & 8\\
 \midrule
 BL2-2s & 48.91 & 67.22 & 78.27 & 0.57 & 2\\
 BL2-8s & 43.46 & 68.38 & 82.84 & 0.55 & 2\\
 \midrule
 BL2-2s + AT & 55.43 & 72.64 & 81.05 & 0.63 & 1\\
 BL2-8s + AT & 66.71 & 84.43 & 91.19 & 0.75 & 1\\
 \bottomrule
\end{tabular}}%
\caption{Retrieval results of attention-based models. 
(R@k = Recall at k, MRR = Mean Reciprocal Rank, MR = Median Rank)}
\label{tab:ismir19_results}
\end{table}


\subsection{Strongly-aligned data constraint}\label{sec:rec_ismir}

In addition to the fixed-size snippet issues discussed above, another limitation of the deep learning 
approach proposed in~\cite{DorferHAFW18_MSMD_TISMIR} relates to its
supervised nature.
In order to generate a large number of matching pairs of short audio and sheet music snippets for training, one 
requires big collections of music data with strong labels (alignment annotations), which 
means fine-detailed mappings between note onsets 
in the audio recordings and their respective note coordinates 
in sheet music images.
Since obtaining such data is labour-consuming and not trivial, the 
embedding learning models rely on synthesised data
(this limitation will be re-visited in the upcoming subsections).

In~\cite{CarvalhoW23_A2S_Recurrent_ISMIR} we propose to address 
both shortcomings in one, by designing a recurrent network that is 
can learn compact and fixed-sized embeddings from longer and 
variable-length passages of audio and sheet music.
The key motivation for this is twofold: by operating with 
variable-length passages, the cross-modal pairs can span the same
music content leading to more robust representations; and by allowing
longer excerpts, we could relax the required annotations from strong  
to weak labels, meaning that now only the corresponding passage boundaries
are needed.
We performed quantitative and qualitative experiments 
in diverse retrieval scenarios with artificial and real data, with 
the results indicating a superior performance of the 
recurrent architectures over the purely convolutional baseline.

\subsection{Generalisation to real-world (noisy) data}
\label{sec:mmsys}

As already hinted at above,
obtaining training data in the form of audio--sheet music datasets with appropriate 
fine-grained alignments is tedious and time-consuming, and also 
requires specialised annotators with proper musical training.
As a consequence, the embedding learning approaches rely on synthetic music
data generated from the Multi-Modal Sheet Music Dataset 
(MSMD)~\cite{DorferHAFW18_MSMD_TISMIR}.
This is a collection of classical piano pieces with rich and varied
data, including score sheets (PDF) engraved via 
Lilypond\footnote{https://www.lilypond.org} and respective
audio recordings synthetised from MIDI with several types of piano 
soundfonts.
With over 400 pieces from several renowned composers, including Mozart, 
Beethoven and Schubert, and covering more than 15 hours of audio, the 
MSMD has detailed audio--sheet music alignments allowing us to obtain 
perfectly matching audio--sheet snippet pairs.
On the downside, the generated scores and audios completely lack real-world artefacts such as scan inaccuracies or room acoustics, and the audios exhibit perfectly steady tempo and dynamics, which is far from how real-world performances would sound.

Using the synthetic MSMD severely affects the capacity of the model 
from Figure~\ref{fig:main} to generalise to realistic retrieval
scenarios when real music data is presented.
In~\cite{CarvalhoWW23_SelfSupLearning_ASR_ACM-MMSys} we proposed
to alleviate this problem via \textit{self-supervised
contrastive learning}.
Inspired by the SimCLR framework~\cite{ChenKNH20_SimCLR_ICML}, we 
pre-trained each independent convolutional pathway 
(see Figure~\ref{fig:main}b) by contrasting differently augmented 
versions of short snippets of audio or sheet music images.
As a key advantage of this approach, the data required for the 
pre-training step needs no annotations, which means we can use 
real music data scraped from the Web.

We applied self-supervised contrastive pre-training for both modalities,
taking the following steps:
\begin{enumerate}
\item Given a sample $\mathbf{x}$ from the training mini-batch of a given
    modality, two stochastic sets of data augmentation transforms are 
    applied to $\mathbf{x}$, generating the positive pair 
    $\tilde{\mathbf{x}}_i$ and $\tilde{\mathbf{x}}_j$.
\item Then a network composed of a CNN encoder and a multi-layer 
    perceptron head computes a latent representation 
    $ \mathbf{z}_i = e(\tilde{\mathbf{x}}_i) $ for each
    augmented sample.
\item Then the normalized-temperature cross-entropy
    (\textit{NT-Xent}) loss 
    function
    is applied and summed over all positive augmented pairs 
    $(\tilde{\mathbf{x}}_i,\tilde{\mathbf{x}}_j)$ within 
    the mini-batch:
\begin{equation}
    \mathcal{L} = 
    \sum_{(i,j)} \mathrm{log}
    \frac{\mathrm{exp}(\mathrm{sim}(z_i, z_j) / \tau)}
    {\sum_{v=1}^{2N}\mathds{1}_{[v \neq i]}
    \mathrm{exp}(\mathrm{sim}(z_i, z_v) / \tau)} \mathrm{,}
\end{equation}\label{eq:nt_xent}
where $\mathrm{sim}(\cdot)$ is the cosine similarity between $z_i$ and 
$z_j$ and the temperature parameter $\tau \in \mathbb{R}_{+}$ 
is adjusted to prioritise poorly embedded snippet pairs.
\end{enumerate}

\begin{figure}[tbp]
  \centering
  \includegraphics[width=0.85\columnwidth]{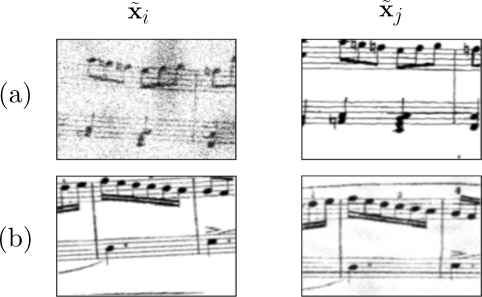}
  \caption{Examples of data augmentation.}
\label{fig:data_aug}
\end{figure}

As for the augmentations used for pre-training, 
we applied to the snippets: horizontal and vertical shifts, resizing 
and rotation, additive Gaussian and Perlin noises, and small and large
elastic deformations.
Figure~\ref{fig:data_aug} shows examples of two pairs of augmented sheet music snippets
when applying all transforms randomly.
The augmentations used on the audio excerpts are: time shift, polarity 
inversion, additive Gaussian noise and gain change, time and frequency
masking, time stretching, and a 7-band equaliser.

In order to investigate the effect of self-supervised contrastive
pre-training on generalisation from synthetic to real data, we prepare
three evaluation datasets: (I) a fully artificial one, using the MSMD; 
(II) a partially real one combining the MSMD with scans of real sheet music; and (III) an entirely real set with audio recordings and 
scanned sheet music images.
We conduct experiments on snippet retrieval as in 
Section~\ref{sec:rec_ismir}, in both search directions 
(audio-to-sheet and sheet-to-audio) 
and compare the baseline model purely trained with MSMD 
as in~\cite{DorferHAFW18_MSMD_TISMIR}, with fine-tuned versions of it
when the audio and sheet music convolutional pathways were pre-trained.

To pre-train the individual pathways, we used raw data acquired from the 
web and public datasets: 200 hours of piano recordings were obtained 
from the MAESTRO dataset~\cite{HawthorneSR+_MAESTRO_ICLR} and 7,000
pages of sheet music were collected from the IMSLP online 
platform\footnote{https://imslp.org}.

Table~\ref{tab:mmsys} provides a summary of the main snippet retrieval 
results in the audio-to-sheet direction.
The baseline model (BL) is compared to its fine-tuned version when both
audio and sheet music CNN pathways were pre-trained using self-supervised
contrastive learning (BL+A+S).
The pre-trained models outperform the baseline in all 
scenarios for all evaluation metrics.
In~\cite{CarvalhoWW23_SelfSupLearning_ASR_ACM-MMSys} the same trend is reported
for the sheet-to-audio direction, indicating that self-supervised pre-training 
is beneficial in our retrieval task.
We note however that there is still a substantial degradation when going from synthetic to partly and, in particular, fully real data. The MRR and MR values in the fully real scenario are definitely still unacceptable for real-world use.

In~\cite{CarvalhoWW23_SelfSupLearning_ASR_ACM-MMSys}, we also evaluated the models on the task of cross-modal \textit{piece identification},
by aggregating snippet embeddings, and also observed better identification results (e.g. over 100\% improvement of the MRR on the task of identifying a piece from an arbitrary recording)
when using the pre-trained models.

\begin{table}
\centering
  \label{tab:mmsys}
  \scalebox{0.92}{
  \begin{tabular}{lcccc}
\toprule
 & \bfseries R@1 & \bfseries R@25 & \bfseries MRR & \bfseries MR \\
\midrule
\multicolumn{5}{l}{(I) \ \ \  MSMD (Fully synthetic data)} \\
\midrule
BL & 0.54 & 0.91 & 0.653 & 1 \\
BL+A+S & 0.57 & 0.93 & 0.687 & 1 \\
\midrule
\multicolumn{5}{l}{(II) \ \ Partially real data} \\
\midrule
BL & 0.28 & 0.67 & 0.375 & 7 \\
BL+A+S & 0.37 & 0.79 & 0.481 & 3 \\
\midrule
\multicolumn{5}{l}{(III) \ Fully real data} \\
\midrule
BL & 0.10 & 0.36 & 0.156 & 76 \\
BL+A+S & 0.15 & 0.48 & 0.226 & 29 \\
\bottomrule
\end{tabular}}
\caption{Audio-to-sheet snippet retrieval results on three types 
  of datasets: (I) fully synthetic, (II) partially real and 
  (III) entirely real.}
\end{table}

\subsection{Temporal dependencies between subsequent snippets}
\label{subsec:eusipco}

As briefly mentioned at the end of Section~\ref{sec:mmsys} and implied in the previous paragraph, a popular task
scenario in the audio--sheet music retrieval realm is \textit{cross-modal piece identification}: 
given an unknown music document in one modality (i.e., a full 
audio recording), we wish to identify which piece is it based 
on a collection of documents in another modality (i.e., a database 
of scanned sheet images).
For deep learning-based embedding methods like
in~\cite{DorferHAFW18_MSMD_TISMIR}, choosing how to aggregate snippet 
embeddings extracted from full documents is essential in order to achieve 
robust piece identification.

The basic identification method proposed 
in~\cite{DorferHAFW18_MSMD_TISMIR} is as follows.
Taking the audio-to-sheet search direction without loss of generality, 
let $ \mathcal{D} $ be a collection of $ L $
sheet music documents, and $ Q $ an unknown audio query.
%
Each document $ D_i \in \mathcal{D} $ is segmented into a 
set of image snippets, which are embedded using the sheet music
pathway of Figure~\ref{fig:main}b, generating a set of sheet music 
embeddings $ \{ y^i_1, y^i_2, ..., y^i_{M_i} \} $ for each piece.
Analogously, the full audio query is segmented into short spectrogram
excerpts, from which a set of query audio embeddings 
$ \{ x_1, x_2, ..., x_N \} $ is computed.
Then for each audio snippet query $ x_j $, its nearest neighbour 
among all embedded image snippets is selected via cosine distance.
Each retrieved sheet snippet then votes for the piece it originated from,
resulting in a ranked list of piece candidates.

A limitation of this vote-based identification procedure is that 
it completely ignores the temporal relationships between subsequent 
snippet queries, which are inherent in, and constitutive of, music.
In~\cite{CarvalhoW21_DTWpieceID_EUSIPCO}, a 
matching strategy is presented that aligns the sequences of embeddings 
obtained from the query document and search database items.
The sequence of embedded snippets $ \{ y^i_1, y^i_2, ..., y^i_{M_i} \} $ 
of each piece $ D_i \in \mathcal{D} $ from the database 
is aligned  to the query sequence $ \{ x_1, x_2, ..., x_N \} $ via 
dynamic time warping (DTW), using the cosine 
distance as a cost function. The DTW alignment cost between 
query $ Q $ and piece $ D_i $ is regarded as the matching cost 
$ c_i = \mathrm{DTW} (Q, D_i) $.
Then a ranked list is computed based on the matching cost of 
each piece to the query, with the best matching piece 
having the lowest alignment cost.

Experiments with real and noisy music data reported 
in~\cite{CarvalhoW21_DTWpieceID_EUSIPCO} reveal that using the 
proposed DTW-based matching strategy improves 
identification results by a large margin, when comparing with 
the simple vote-based approach.
%
However a number of additional shortcomings of this proposed
matching strategy arise: first, even though there are fast 
implementations of the DTW algorithm, the retrieval time scales up 
considerably as the search database grows. Moreover, DTW 
does not handle typical structural differences between audio 
performances and scores, caused by, e.g., repeats that are, or are not, played.
Therefore we believe the next steps in this direction should
target algorithms that can scale to large music collections,
in terms of processing time, and that are flexible and robust in dealing 
with structural mismatches between audio and sheet music.

\Section{Remaining Challenges}
In this section, we briefly discuss the remaining obstacles and open 
problems, and identify promising directions for future research.

\subsection{Public large-scale datasets}
Large and licence-free datasets are invaluable resources in audio--sheet 
music retrieval research, enabling the training of deep learning models, 
facilitating benchmarking and comparative studies, promoting reproducibility, 
encouraging innovation, and ensuring the relevance of developed methods to 
real-world applications.

For the audio part, existing public datasets such as 
MAESTRO~\cite{HawthorneSR+_MAESTRO_ICLR} provide a considerable number
of piano audio recordings.
However when targeting truly large-scale databases, 
Youtube\footnote{https://www.youtube.com} can be a 
valuable source for collecting audio recordings, via using its API or
scrapping techniques.
Together with big amounts of curated sheet music PDFs obtained from 
online libraries like IMSLP, researchers can create large-scale audio-sheet music datasets that enable the development and evaluation of robust retrieval methodologies.
However, it is important to ensure compliance with copyright laws, respect data usage policies of the platforms involved, and provide appropriate attribution when using data from third-party sources.

\subsection{Efficient structures for fast retrieval}

Quick responses are pivotal in audio-sheet music retrieval 
research, particularly in large-scale scenarios, as they enhance 
efficiency, improve the user experience, ensure scalability, enable 
practical deployment, and support real-time feedback and iterative 
refinement. 
%

A potential direction is to use compact cross-modal fingerprints, 
which allow fast search in music as 
in~\cite{ArztBW12_SymbolicFingerprint_ISMIR,
Tsai20_LinkingLakhtoIMSLP_Bootleg_ICASSP}.
Moreover such algorithms should be flexible to handle any kind of 
structural mismatch between an audio performance and a printed score, 
as discussed in Subsection~\ref{subsec:eusipco}.

\subsection{Instrumentation and genre}

Incorporating diverse instrumentation and genres in audio-sheet music 
retrieval research enables the development of more inclusive, 
adaptable, and effective retrieval methods that align with real-world 
scenarios and user expectations.
It broadens the scope of the field 
and promotes advancements that cater to the diverse musical landscape 
found in big and heterogeneous music collections.

Most retrieval methods use classical piano music as a case study since 
this type of data is easier to collect due to its abundance and 
popularity.
Complex types of scores, such as orchestral and jazz music, will 
required more sophisticated and flexible methods, which could be
assisted for example by optical music 
recognition~\cite{Calvo-ZaragozaHP21_OMRReview_ACM}.

\Section{Conclusion}

This article examined the current developments in audio-sheet music 
retrieval via deep learning methods. 
We have identified the main obstacles on the road towards robust and 
large-scale retrieval and have discussed the steps 
taken to address some of these challenges. 
While there has been steady progress in the field over the past years,
there are still open problems that hinder
the large-scale employment of this methodology.

To assist the progress towards a unified and robust retrieval 
methodology for cross-modal music retrieval, we believe it is 
crucial to address these remaining challenges, this way 
unlocking new possibilities for connecting large and heterogeneous 
music collections and contribute to the enrichment of music 
information retrieval applications.

\section*{Acknowledgments}

This work is supported by the European Research Council
(ERC) under the EU’s Horizon 2020 research and innovation
programme, grant agreement No.~101019375 (\textit{Whither Music?}), and the Federal State of Upper Austria (LIT AI Lab).

\bibliographystyle{latex8}
\bibliography{latex8}

\end{document}